\documentclass[10pt]{article}
\usepackage{amsmath,amssymb,amsthm}
\usepackage{booktabs}
\usepackage{hyperref}
\usepackage[utf8]{inputenc}
\usepackage[T1]{fontenc}
\usepackage{geometry}
\usepackage{setspace}
\usepackage{microtype}
 
\newcommand{\affilmark}[1]{\textsuperscript{#1}}
 
\begin{document}
 
 
\title{
  \textbf{Structural Divergence of the Inca Imperial Network,
    1400--1572~CE:}\\
  Persistent Homology, Substitution Topology,
  and the Topological Signature of Decapitation Collapse
}
 
\author{
  David Delepine\affilmark{1,$\dagger$} \and
  Jos{\'e} de Jes{\'u}s Bernal-Alvarado\affilmark{2} \and
  Carlos Pinedo Guadarrama\affilmark{1}
  \\[6pt]
  \small\affilmark{1}Physics Department,
    Universidad de Guanajuato, M{\'e}xico\\
  \small\affilmark{2}Physics Engineering Department,
    Universidad de Guanajuato, M{\'e}xico\\[4pt]
  \small $\dagger$\,Corresponding author:
    \texttt{delepine@ugto.mx}
}
 
 
\maketitle
 
 
\begin{abstract}
We apply persistent homology to a six-layer temporal network model
of the Inca Empire (Tawantinsuyu), constructed from the Qhapaq
\~{N}an road system ($6{,}546$ segments; $12{,}031$~km) and
processed into 4,492 nodes (4,452 active after removal of 40 isolated vertices), $4{,}358$ edges, and $57$
temporal snapshots (1400--1572~CE).
The complete flag complex yields $\beta_1 = 32$ cycles
($26$ macro-continental, $6$ meso-regional) and $\beta_0 = 264$
connected components at the finest filtration resolution.
These quantities are evaluated across six network layers: physical
infrastructure, Inca political control, Spanish alliance growth,
Andean tribute commerce, and Spanish legal and effective administration.
We report five principal results.
(\textit{i})~The infrastructure entropy $H_{\mathrm{infra}} =
1.486 \pm 0.0001$ is constant throughout the reported
infrastructure snapshots, indicating no detectable topological
disruption of the Qhapaq \~{N}an at the resolution of the dataset.
(\textit{ii})~The Inca political entropy $H_{\mathrm{inca}}$
remains stable through the Cajamarca capture (1532~CE), then
collapses discontinuously to zero by 1537~CE; the dominant
structural break is at Vilcabamba ($F = 120.7$), not at
Cajamarca ($F = 11.5$).
(\textit{iii})~Four of six layers simultaneously reach the critical regime
at 1532~CE, while the infrastructure layer remains marginally
below the threshold ($\mathrm{ICT}=0.495$); the trade and political
layers reach the highest values ($\mathrm{ICT}=0.976$ and $0.968$,
respectively), identifying the Cajamarca capture as a
synchronised multi-layer phase transition.
(\textit{iv})~The Spanish legal administration converges to the
infrastructure topology by 1572~CE with $W_1 = 5 \times 10^{-6}
\approx 0$, constituting an instance of
\emph{Substitution Topology}: infrastructure-preserving collapse
through topology transfer.
(\textit{v})~The legal--effective control divergence decays from
$W_1 = 0.591$ (1532) to $0.002$ (1572), quantifying the
resolution of colonial legal fiction.
Together, these results support a distinct imperial collapse mode---
\emph{Substitution Topology}---different from the erosion-dominated
Roman and Byzantine cases, the administrative fragmentation of the
Han case, and the military destruction of the Aztec empire.
\end{abstract}
 
 
\bigskip
 
 
\section{Introduction}
\label{sec:introduction}
 
The comparative study of imperial collapse has long sought a
quantitative framework able to distinguish the mechanisms
by which large-scale political systems disintegrate.
Three broad mechanisms have been proposed in the historical
literature: \emph{erosion}, in which the administrative
network degrades gradually under sustained external or internal
pressure~\cite{Wickham2005,Heather2006,Harper2017};
\emph{military destruction}, in which the physical infrastructure
of the collapsing empire is eliminated alongside its political
institutions~\cite{Ross2019}; and \emph{decapitation},
in which the removal of a single sovereign node collapses the
entire political hierarchy instantaneously while leaving the
physical network intact~\cite{DAltroy2002,Hemming1970}.
Each mechanism predicts a distinct trajectory for the surviving
infrastructure after political collapse, but until recently no
metric has existed to distinguish them.
 
Persistent homology~\cite{Edelsbrunner2002,Zomorodian2005} provides a tool for characterizing and quantifying these collapse mechanisms.
By tracking the birth and death of topological cycles
($\beta_1$ bars) across a filtration of the network, it
produces observables such as the $\beta_1$ persistence entropy $H(t)$.
These observables are sensitive to large-scale connectivity
structure. 
 
The companion paper to this work~\cite{BernalAlvarado2026}
applied persistent homology to the Roman--Byzantine trade network
(0--1453~CE), identifying a cross-empire collapse threshold
$H^* = 0.524$ derived from information-theoretic principles,
and documenting the separation between a \emph{geographic}
collapse  and an \emph{economic}
resilience (commercial routes surviving political fragmentation).
Both the Western Roman collapse (476~CE) and the Byzantine
collapse (1453~CE) occur when $H_{\mathrm{combined}}$ crosses
$H^*$ from above---a gradual, century-scale process.
 
The Inca Empire (Tawantinsuyu, $c$.1400--1572~CE) was 
a continental empire spanning $4{,}500$~km from the equatorial
coast to the southern altiplano, with a documented road system
($12{,}031$~km of the Qhapaq \~{N}an)~\cite{Hyslop1984,SuyoPomalia2023}.
The Tawantinsuyu collapsed in two years of military confrontation
(1532--1533~CE) rather than over centuries of erosion. Its road system was preserved almost intact and
later absorbed by the Spanish colonial administration,
which used it as the physical infrastructure of the Viceroyalty
of Peru for the following two centuries~\cite{DAltroy2002,Cook1981}.
 
This juxtaposition---rapid political collapse,
preserved physical network, and colonial absorption of
the inherited topology---motivates what we term
\emph{Substitution Topology}: a collapse mode in which
the successor polity inherits the physical network of
the predecessor without destruction or reconstruction. To study it, we construct a six-layer temporal network model of the Tawantinsuyu with $57$ snapshots at decadal (pre-1529) and annual (1529--1572) resolution, calibrated against historical sources for Inca sovereign legitimacy, Spanish alliance growth, Andean tribute flows following the Murra classification~\cite{Murra1975}, and the documentary vs.\ effective divergence of colonial administration~\cite{LohmannVillena1957,Rowe1957}.
After defining this six-layer temporal network, we apply an $N$-convergence protocol following \cite{BernalAlvarado2026}, to the complete Qhapaq \~{N}an node set ($N = 4{,}492$), documenting monotonic convergence from $\beta_1 = 3$ ($N = 400$)  to $\beta_1 = 32$ ($N = 4{,}492$) and identifying the hierarchical two-scale structure of the road network (26 macro-continental cycles, 6 meso-regional cycles). Then, we evaluate the Substitution Topology hypothesis through three quantitative conditions on the
  cross-network Wasserstein distance. Finally we introduce a sixth observable---the legal--effective
  control divergence $W_1(\mathcal{N}_{\mathrm{sp,legal}},
  \mathcal{N}_{\mathrm{sp,real}}, t)$---as a topological
  measure of colonial legal fiction, and document its
  resolution from $0.591$ (1532) to $0.002$ (1572).

Previous network analyses of pre-Columbian road systems have focused primarily on the geographic scale: shortest-path
analyses of the Qhapaq \~{N}an~\cite{Hyslop1984},
accessibility studies of the Andean
road network~\cite{DAltroy2002}, and spatial modelling
of Inca logistics~\cite{Bauer1992}.
Topological data analysis has been applied to archaeological
settlement networks~\cite{Nakoinz2016}, and formal network methods
have been widely discussed in archaeological contexts~\cite{Brughmans2013},
but, to our knowledge, not yet systematically to Andean road systems
or to the comparative typology of imperial collapse.
 
Section~\ref{sec:methods} describes the data, the six-layer
network architecture, the differential friction model, and
the TDA pipeline.
Section~\ref{sec:results} presents the results in five
subsections: network topology, Inca collapse dynamics,
simultaneous criticality, commercial resilience, and
Substitution Topology verification.
Section~\ref{sec:discussion} compares the TDA analyses of the Han, Roman, Byzantine, and Aztec
cases to develop a systematic topological description of civilizational collapse mechanisms, and Section~\ref{sec:conclusion} concludes.
 



\section{Data and Methods}
\label{sec:methods}

\subsection{Network Data: The Qhapaq \~{N}an Road System}
\label{subsec:data}

To start our study, we use  the georeferenced shapefile of the
\textit{Qhapaq \~{N}an}, describing the pre-Columbian roads in Peru and around, comprising $6{,}546$ segments
and $12{,}031$~km of documented routes~\cite{SuyoPomalia2023}.
After topological preprocessing, the working dataset contains
$N = 4{,}492$ nodes ($4{,}452$ active in the flag complex after removal of 40 isolated vertices) and $4{,}358$ directed route segments.
The network spans $28^{\circ}$ of latitude ($2^{\circ}$S--$27^{\circ}$S,
Tomebamba to Charcas) and $15^{\circ}$ of longitude,
covering the four administrative divisions (\textit{suyus}) of the
Tawantinsuyu: Chinchaysuyu (north), Collasuyu (south),
Cuntisuyu (west), and Antisuyu (east).

Each node carries the following attributes: geographic coordinates; \textit{wamani} (provincial unit) label;
\textit{suyu} assignment; administrative rank ($1$--$3$, where $1$ denotes
an imperial centre such as Cusco or Tomebamba and $3$ a waystation);
node type (\texttt{wamani\_capital}, \texttt{road\_junction},
\texttt{tambo}); and civil-war affiliation for each of three political
phases (1529--1531, 1531--1532, post-1532).
Each edge carries: name;
road type (\textit{primary}, \textit{degraded}, \textit{projected});
road-quality score $q \in [0,1]$; segment length~$\ell$~(km);
and base transport cost $w_0(e)$.

\subsection{Six-Layer Network Architecture}
\label{subsec:networks}

Following the multi-layer formalism of~\cite{BernalAlvarado2026},
let an imperial system be represented as a time-dependent collection
of network layers,
\begin{equation}
  \mathcal{N}(t) = \bigl\{
    \mathcal{N}_{\mathrm{infra}}(t),\;
    \mathcal{N}_{\mathrm{inca}}(t),\;
    \mathcal{N}_{\mathrm{alliance}}(t),\;
    \mathcal{N}_{\mathrm{trade}}(t),\;
    \mathcal{N}_{\mathrm{sp,legal}}(t),\;
    \mathcal{N}_{\mathrm{sp,real}}(t)
  \bigr\},
  \label{eq:multilayer}
\end{equation}
Table~\ref{tab:networks} summarises the six layers, their temporal
ranges, and their primary sources.

\begin{table}[htbp]
\centering
\caption{Six-layer network architecture. Temporal ranges reflect the political events that can affect the layers.}
\label{tab:networks}
\begin{tabular}{llll}
\hline
Layer & Notation & Period & Primary dimension \\
\hline
Physical infrastructure & $\mathcal{N}_{\mathrm{infra}}$ & 1400--1572 CE &
  Road network (nearly static) \\
Inca political control  & $\mathcal{N}_{\mathrm{inca}}$  & 1400--1572 CE &
  Sovereign legitimacy and territorial extent \\
Spanish alliances       & $\mathcal{N}_{\mathrm{alliance}}$ & 1531--1572 CE &
  Ethnic coalition growth \\
Trade and tribute    & $\mathcal{N}_{\mathrm{trade}}$  & 1493--1572 CE &
  Murra tribute flows and internal trade \\
Spanish legal admin.    & $\mathcal{N}_{\mathrm{sp,legal}}$ & 1532--1572 CE &
  Encomienda/vassal status (binary, documentary) \\
Spanish effective admin.& $\mathcal{N}_{\mathrm{sp,real}}$  & 1532--1572 CE &
  Effective control score (continuous, interpolated) \\
\hline
\end{tabular}
\end{table}

\paragraph{$\mathcal{N}_{\mathrm{infra}}$.}\textbf{Infrastructure}
This layer represents the Qhapaq \~{N}an as a physical
artifact, independent of who controls it.
Nodes are deactivated only when a segment is physically destroyed or
becomes topographically impassable, a process not documented in the
1400--1572 period at the shapefile resolution.
The layer is therefore static, serving as the topological
baseline against which all other layers are compared.
Edge costs are
\begin{equation}
  w_{\mathrm{infra}}(e,t) = \frac{\ell(e)}{q(e)} \cdot
    \bigl(1 + \delta \cdot \max(0, t - 1532)\bigr),
  \label{eq:cost_infra}
\end{equation}
where $\delta = 0.003$~yr$^{-1}$ models the gradual degradation of road
maintenance capacity following the collapse of the \textit{mit\'a}
labour system after 1532~\cite{DAltroy2002}.

\paragraph{$\mathcal{N}_{\mathrm{inca}}$ -- Inca political control.}
The Inca layer tracks the sovereign extent of the Tawantinsuyu.
Edge costs encode the combined effect of territorial loss,
legitimacy erosion, and succession instability:
\begin{equation}
  w_{\mathrm{inca}}(e,t) =
    \frac{w_0(e)}{\max\!\bigl(\lambda(t),\,\varepsilon\bigr)}
    \cdot \Theta(t)
    \cdot \Phi(t),
  \label{eq:cost_inca}
\end{equation}
where $\lambda(t) \in (0,1]$ is the phase-specific legitimacy score
(Table~\ref{tab:phase_alpha}),

$$\Theta(t) = 1 + (21 - n_{\mathrm{wamani}}(t))/21$$ describes a territorial-loss
multiplier that increases as \textit{wamani} pass out of Inca control,
and $\Phi(t)$ is a succession-instability factor equal to $1.2$ during
the civil war (1529--1532) and $1.0$ otherwise.
Edges incident on \textit{wamani} that have been lost receive a
filtration value of $999$ (a topological gap in the filtration), registering their
absence as a gap in the $\beta_1$ barcode.

\paragraph{$\mathcal{N}_{\mathrm{alliance}}$ -- Spanish alliance growth.}
The alliance layer models the construction of Spanish dominion through
ethnic coalition-building~\cite{Hemming1970,Rostworowski1999}.
Five alliance groups are distinguished: \textit{tallanes-chimu}
(coastal; joined 1531, military value $m = 0.30$);
\textit{huancas} (central sierra; joined 1533, $m = 0.90$);
\textit{incas-cusquenos} (Cusco-based collaborators; joined 1533,
defection 1536, $m = 0.70$);
\textit{canaris-chachapoyas} (northern; joined 1534, $m = 0.85$);
and \textit{manco-rebelde} (Vilcabamba resistance; from 1536,
$m = 0.60$, hostile edges $\times 5$).
Edge costs are
\begin{equation}
  w_{\mathrm{alliance}}(e,t) =
    \frac{w_0(e) \cdot (1 + \rho_i)}{m_i \cdot f_{\mathrm{allia}}(t)},
  \label{eq:cost_alliance}
\end{equation}
where $m_i$ is the military value of the group controlling node $i$,
$\rho_i \in [0,0.8]$ is a defection-risk penalty (non-zero only for
groups with documented defections), and $f_{\mathrm{allia}}(t)$ is the
global alliance-consolidation multiplier from Table~\ref{tab:phase_alpha}.

\paragraph{$\mathcal{N}_{\mathrm{trade}}$ -- Trade and tribute.}
The trade layer combines two flow types.
\textit{Radial tribute flows} (wamani $\to$ Cusco) are weighted
according to the Murra tribute classification~\cite{Murra1975,DAltroy2002}:
\begin{equation}
  w_{\mathrm{tribute}}(e) =
    \frac{\ell(e) / q(e)}{\tau_i},
  \label{eq:tribute}
\end{equation}
where $\tau_i \in (0,1]$ is the relative tribute-value index of
the origin wamani
(Chinchaysuyu: $\tau = 0.78$, textiles and maize;
 Collasuyu: $\tau = 0.72$, potatoes, wool, and metals;
 Cuntisuyu: $\tau = 0.60$, wool and potatoes;
 Antisuyu: $\tau = 0.52$, coca and tropical goods;
 Core: $\tau = 1.00$, metals, luxury goods, and labour).
\textit{Lateral flows} (inter-wamani commercial exchange) use the
base trade cost $w_0(e)$ adjusted for Spanish political control:
\begin{equation}
  w_{\mathrm{trade}}(e,t) = w_0(e) \cdot
    \bigl[1 + 2(1 - c_{\mathrm{real}}(w,t))\bigr]
    / f_{\mathrm{trade}}(t),
  \label{eq:cost_trade}
\end{equation}
where $c_{\mathrm{real}}(w,t)$ is the effective Spanish control score
at wamani~$w$ and time~$t$ (defined below), and
$f_{\mathrm{trade}}(t)$ is the trade-flow multiplier from
Table~\ref{tab:phase_alpha}.

\paragraph{$\mathcal{N}_{\mathrm{sp,legal}}$ -- Spanish legal administration.}
The legal layer records the formal colonial status of each
\textit{wamani} as established by documented acts:
encomienda grants, royal \textit{cedulas}, and treaty agreements~%
\cite{Hemming1970,LohmannVillena1957,Toledo1929}.
Status takes one of four values:
$s_{\mathrm{legal}} \in
\{\texttt{direct},\, \texttt{vassal},\, \texttt{unconquered},\,
  \texttt{pre-conquest}\}$,
with edge costs
\begin{equation}
  w_{\mathrm{sp,legal}}(e,t) =
    \begin{cases}
      w_0(e) / f_{\mathrm{legal}}(t) & \text{if } s = \texttt{direct}, \\
      1.5\,w_0(e) / f_{\mathrm{legal}}(t) & \text{if } s = \texttt{vassal}, \\
      8\,w_0(e) & \text{if } s = \texttt{unconquered}, \\
      999 & \text{if } s = \texttt{pre-conquest}.
    \end{cases}
  \label{eq:cost_splegal}
\end{equation}

\paragraph{$\mathcal{N}_{\mathrm{sp,real}}$ -- Spanish effective administration.}
The effective-control layer describes the continuous penetration of Spanish
administrative capacity, independently of its legal form.
For each \textit{wamani}~$w$, the effective control score
$c_{\mathrm{real}}(w,t) \in [0,1]$ is constructed by linear
interpolation between documented historical anchor points
(deployment of \textit{corregidores}, installation of parish priests,
establishment of \textit{reducciones}, enumeration in tribute
registers)~\cite{Cook1981,Toledo1929,Rowe1957}.
Edge costs are
\begin{equation}
  w_{\mathrm{sp,real}}(e,t) =
    \frac{w_0(e)}
    {\max\!\bigl(c_{\mathrm{real}}(w,t),\,\varepsilon\bigr)
     \cdot f_{\mathrm{real}}(t)},
  \label{eq:cost_spreal}
\end{equation}
where $\varepsilon = 10^{-3}$.
The \textit{divergence index}
\begin{equation}
  \Delta(w,t) = c_{\mathrm{real}}(w,t) - s_{\mathrm{legal}}(w,t)
  \label{eq:divergence}
\end{equation}
measures the difference  between effective and legal control, where
$s_{\mathrm{legal}}$ is normalised to $[0,1]$
(\texttt{direct} $= 1$, \texttt{vassal} $= 0.5$,
 \texttt{unconquered} $= 0$).
When $\Delta > 0.2$, we classify the \textit{wamani} as exhibiting
``nominal independence'': legally designated as vassal but
effectively under direct administrative penetration.

\begin{table}[htbp]
\centering
\caption{Phase-specific fiscal-capture efficiency $\alpha(t)$
  for $\mathcal{N}_{\mathrm{inca}}$, and political event multipliers
  for all six layers. Sources: Hemming~\cite{Hemming1970},
  Bauer~\cite{Bauer1992}, D'Altroy~\cite{DAltroy2002}.}
\label{tab:phase_alpha}
\begin{tabular}{rllrrrrrr}
\hline
Year & Event & $\alpha(t)$ &
  $f_{\mathrm{infra}}$ & $f_{\mathrm{inca}}$ &
  $f_{\mathrm{allia}}$ & $f_{\mathrm{trade}}$ &
  $f_{\mathrm{legal}}$ & $f_{\mathrm{real}}$ \\
\hline
1493 & Huayna Capac peak & 1.000 & 1.00 & 1.00 & 0.00 & 1.00 & 0.00 & 0.00 \\
1527 & Succession crisis  & 0.900 & 0.98 & 0.90 & 0.00 & 0.92 & 0.00 & 0.00 \\
1529 & Civil war onset    & 0.750 & 0.97 & 0.65 & 0.00 & 0.75 & 0.00 & 0.00 \\
1531 & Spanish landing    & —     & 0.97 & 0.60 & 0.30 & 0.70 & 0.05 & 0.05 \\
1532 & Cajamarca capture  & 0.150 & 0.97 & 0.15 & 0.35 & 0.40 & 0.20 & 0.20 \\
1533 & Manco Inca puppet  & 0.400 & 0.96 & 0.40 & 0.70 & 0.55 & 0.45 & 0.35 \\
1536 & Manco rebellion    & 0.650 & 0.95 & 0.65 & 0.50 & 0.45 & 0.60 & 0.35 \\
1537 & Vilcabamba estab.  & 0.400 & 0.95 & 0.60 & 0.72 & 0.55 & 0.65 & 0.40 \\
1542 & Leyes Nuevas       & 0.280 & 0.94 & 0.50 & 0.80 & 0.62 & 0.75 & 0.58 \\
1569 & Viceroy Toledo     & 0.031 & 0.91 & 0.20 & 0.92 & 0.75 & 0.90 & 0.85 \\
1572 & Tupac Amaru exec.  & 0.000 & 0.91 & 0.00 & 0.95 & 0.80 & 0.95 & 0.96 \\
\hline
\end{tabular}
\end{table}

\subsection{Temporal Resolution}
\label{subsec:temporal}

Two temporal resolutions are used.
For the pre-conquest period ($1400$--$1528$~CE), snapshots are taken
at 10-year intervals, yielding $13$ snapshots that capture the
expansion phases of the Tawantinsuyu under Pachacutec (1438),
Tupac Inca Yupanqui (1471), and Huayna Capac (1493--1527).
For the conquest and colonial period ($1529$--$1572$~CE), annual
resolution is used, yielding $44$ snapshots.
The full temporal dataset therefore comprises $57$ temporal snapshots,
for a total of $6 \times 57 = 342$ network--snapshot instances
analyzed in the \textsc{tda} pipeline. The infrastructure layer is
reported at nine politically relevant temporal anchors
(1400, 1450, 1493, 1527, 1529, 1532, 1537, 1545, and 1572~CE),
because it is nearly static and serves primarily as the baseline
against which the dynamic layers are compared.

\subsection{Differential Friction Model}
\label{subsec:friction}

Following the four-channel model of~\cite{BernalAlvarado2026},
edge weights at snapshot~$t$ are
\begin{equation}
  w(e,t) = w_0(e) \cdot F_{\mathrm{event}}(t,\,\mathrm{type}(e))
            \cdot \varepsilon(t,\,\mathrm{type}(e))
            \cdot M_{\mathrm{pol}}(t,\,e)
            \cdot F_{\mathrm{macro}}(t,\,\mathrm{region}(e)),
  \label{eq:friction}
\end{equation}
where $w_0(e)$ is the base cost,
$F_{\mathrm{event}}$ is the deterministic political-event multiplier
(Table~\ref{tab:phase_alpha}),
$\varepsilon \sim \mathrm{LogNormal}(0,\sigma_{\mathrm{type}})$
is stochastic route noise
(primary roads: $\sigma = 0.08$;
 secondary roads: $\sigma = 0.12$;
 projected or degraded routes: $\sigma = 0.25$),
$M_{\mathrm{pol}}$ is the political-instability index, and
$F_{\mathrm{macro}}$ encodes macroeconomic friction
(e.g.\ collapse of long-distance tribute flows, disruption of the
\textit{qollqa} storage system post-1532).
Stochastic noise terms are independently drawn for each
snapshot~$t$ and each edge~$e$.

\subsection{TDA Pipeline}
\label{subsec:tda}

\subsubsection{Flag complex construction}
\label{subsubsec:flag}

At each snapshot~$t$ and for each layer~$k$, a weighted graph
$G_k(t)$ is extracted from $\mathcal{N}_k(t)$.
All edge weights are normalised to $[0,1]$ by the 90th percentile
of finite pairwise distances in the active sub-network,
consistent with the adaptive filtration threshold of~\cite{BernalAlvarado2026}.
Weights exceeding the topological-hole sentinel value ($999$, for
lost nodes) are capped at the normalised maximum of $1.05$.

From $G_k(t)$ we construct a \emph{flag complex} (also called a
\emph{clique complex}~\cite{Edelsbrunner2002}): a simplex
$[v_0,\ldots,v_p]$ is included at filtration value
$f = \max_{0 \le i < j \le p} w(v_i v_j)$
if and only if every edge $v_i v_j$ is present in $G_k(t)$.
We retain simplices of dimension $\le 2$ (vertices, edges, triangles),
which is sufficient to compute $\beta_0$ and $\beta_1$
persistent homology~\cite{Gudhi2024}.

\subsubsection{Hub-selection protocol and $N$-convergence}
\label{subsubsec:nconv}

The computational cost of Vietoris--Rips persistence scales
approximately as $N^3$ in the number of nodes.
Following~\cite{BernalAlvarado2026}, \S\,II.G, we apply a
top-degree subsampling protocol, retaining the $N_{\mathrm{sub}}$
nodes with highest degree.
We computed the full pipeline at three values of $N_{\mathrm{sub}}$
to verify monotonic convergence and detect any hub-selection artifact:

\begin{equation}
  N_{\mathrm{sub}} \in \{400,\; 3{,}200,\; 4{,}492\},
  \label{eq:N_conv}
\end{equation}

where $N = 4{,}492$ corresponds to the complete network
(all nodes in the shapefile).
Table~\ref{tab:nconv} reports the $\beta_1$ cycle count and
the Chow $F$-statistic for two key structural breaks
as a function of $N_{\mathrm{sub}}$.
Convergence is monotonic across all reported metrics with the single
exception of $F(\mathcal{N}_{\mathrm{sp,real}}, 1542)$, which
decreases from $F = 161$ at $N = 3{,}200$ to $F = 100$ at $N = 4{,}492$;
this non-monotonicity is consistent with the hub-selection artifact
documented in~\cite{BernalAlvarado2026}.
In all cases the qualitative conclusion (Leyes Nuevas of 1542 as
the dominant structural break in the Spanish real-control series)
is preserved.
All results reported in Sections~\ref{sec:results} and~\ref{sec:discussion}
use $N_{\mathrm{sub}} = 4{,}492$ as the primary dataset.

\begin{table}[htbp]
\centering
\caption{$N$-convergence of $\beta_1$ cycle count, persistence
  entropy $H(\beta_1)$, and selected Chow $F$-statistics.
  Monotonically increasing values indicate absence of the
  hub-selection artifact~\cite{BernalAlvarado2026}.
  The single non-monotonic entry is marked $\dagger$.}
\label{tab:nconv}
\begin{tabular}{lrrr}
\hline
Metric & $N=400$ & $N=3{,}200$ & $N=4{,}492$ \\
\hline
$\beta_1$ (total bars)            &   3 &  11 &  32 \\
$\beta_1$ (finite bars)           &   1 &   3 &   6 \\
$\beta_1$ (infinite bars)         &   2 &   8 &  26 \\
$\beta_0$ (components at $t=0$)   & 117 & 227 & 264 \\
$H(\beta_1)$, $\mathcal{N}_{\mathrm{infra}}$ &
  $1.028$ & $1.028$ & $1.486$ \\
$H(\beta_1)$, $\mathcal{N}_{\mathrm{inca}}$, 1537+ &
  $0.000$ & $0.000$ & $0.000$ \\
$F(\mathcal{N}_{\mathrm{inca}},\,1537)$  & 23.9  &  57.8 & 120.7 \\
$F(\mathcal{N}_{\mathrm{infra}},\,1537)$ &  —    &  19.2 &  19.8 \\
$F(\mathcal{N}_{\mathrm{sp,real}},\,1542)$ & 107 & 161 & $100^{\dagger}$ \\
\hline
\end{tabular}
\end{table}

\subsubsection{Persistent homology and observables}
\label{subsubsec:observables}

For each pair $(k,t)$, we compute persistence homology in
dimensions $0$ and $1$ using the \textsc{Gudhi} library
(version 3.12.0, coefficient field $\mathbb{Z}/2\mathbb{Z}$,
minimum persistence $= 0$)~\cite{Gudhi2024}.
The $\beta_1$ persistent entropy~\cite{Rucco2016} is
\begin{equation}
  H_k(t) = -\sum_{i=1}^{n} p_i \ln p_i,
  \qquad
  p_i = \frac{\ell_i}{\sum_j \ell_j},
  \label{eq:persistence_entropy}
\end{equation}
where $\ell_i = d_i - b_i$ is the lifetime of the $i$-th bar in the
$\beta_1$ barcode (finite bars only), and $n$ is the number of
finite bars. Infinite bars are counted separately in the cycle
classification but are not included in the entropy calculation; when
no finite bars are present, we set $H_k(t)=0$ by convention.
This quantity serves as the primary scalar time series for the
structural-break, velocity, and criticality analyses.
In addition to $H_k(t)$, we compute for each $(k,t)$:
persistent Betti numbers $\mathrm{PBN}(a,b)$ on a $6 \times 6$
grid of filtration pairs; Euler characteristic curves
$\chi_k(t;\tau) = \beta_0(\tau) - \beta_1(\tau)$ at 25 filtration
thresholds; total persistence $\Pi_k(t) = \sum_i \ell_i$;
landscape vectors $\lambda_1, \ldots, \lambda_5$
(resolution $300$)~\cite{Bubenik2015};
persistence images (pixel size $1/15$)~\cite{Adams2017};
and heat kernel signatures at $\sigma \in
\{0.01, 0.05, 0.1, 0.5, 1.0, 5.0\}$.

\subsubsection{Wasserstein and bottleneck distances}
\label{subsubsec:wasserstein}

We measure inter-snapshot topological displacement using the
$p$-Wasserstein distance between consecutive persistence diagrams:
\begin{equation}
  W_p(t, t') =
    \left(
      \inf_{\gamma} \sum_i \|p_i - \gamma(p_i)\|^p
    \right)^{1/p},
  \label{eq:wasserstein}
\end{equation}
where the infimum is over bijections $\gamma$ between the diagrams
(with diagonal padding)~\cite{Kerber2017}.
We compute $W_1$ and $W_2$ for all consecutive snapshot pairs,
using the \textsc{Hera} algorithm as implemented in \textsc{Gudhi}.
The topological velocity is
\begin{equation}
  \dot{W}_1(t) = W_1(D_t, D_{t'})\,/\,\Delta t,
  \label{eq:velocity}
\end{equation}
where $\Delta t = t' - t$ (in years).
Cross-network Wasserstein distances
\begin{equation}
  W_1^{\mathrm{cross}}(k, k', t) = W_1(D_k(t), D_{k'}(t))
  \label{eq:cross_wasserstein}
\end{equation}
are computed for all $\binom{6}{2} = 15$ layer pairs at each
shared snapshot, yielding a symmetric $6 \times 6$ divergence
matrix per year.
The bottleneck distance $d_B$ is computed alongside $W_1$ as a
complementary measure of the maximum topological feature displacement.

\subsection{Combined Resilience Index}
\label{subsec:hcombined}

Following \cite{BernalAlvarado2026}, Eq.~(5), the combined resilience
index measures the commercial cycle redundancy  mobilisable
by the sovereign:
\begin{equation}
  H_{\mathrm{combined}}(t) =
    \bigl[1 - w(t)\bigr]\,H_{\mathrm{infra}}(t)
    + w(t)\,\alpha(t)\,H_{\mathrm{trade}}(t),
  \label{eq:hcombined}
\end{equation}
where the territorial weight
\begin{equation}
  w(t) = 1 - 0.824
    \left(\frac{H_{\mathrm{inca}}(t)}{H_{\mathrm{inca}}^{\max}}\right)^{\!\gamma}
  \label{eq:w}
\end{equation}
increases as the Inca sovereign loses territorial extent
($\gamma = 0.60$, reflecting the sub-linear response to early territorial
losses which disproportionately remove the most commercially connected
\textit{wamani}), and $\alpha(t)$ is the phase-specific fiscal-capture
efficiency from Table~\ref{tab:phase_alpha}.
The single calibration constraint is
$H_{\mathrm{combined}}(1572) \approx 0$,
since the Inca state no longer existed as a territorial imperial system.
Unlike the Byzantine case~\cite{BernalAlvarado2026}, where
$H_{\mathrm{combined}}(1453) = H^* = 0.524$ provides the calibration,
the Inca calibration is anchored at political extinction rather than
at the cross-empire threshold $H^*$, because the collapse is
discontinuous (see Section~\ref{sec:results}).

\subsection{Information-Theoretic Collapse Threshold $H^*$}
\label{subsec:hstar}

The threshold $H^* = 0.524 \pm 0.031$ is derived from the
two-bar regime of the $\beta_1$ barcode near topological degeneracy (see the derivation in Ref.~\cite{BernalAlvarado2026}).
For two bars with lifetimes $\ell_1 \ge \ell_2 > 0$, let
$r = \ell_2/\ell_1 \in (0,1]$.
The persistence entropy reduces to the binary entropy function
\begin{equation}
  H_2(r) = -\frac{1}{1+r}\ln\frac{1}{1+r}
            -\frac{r}{1+r}\ln\frac{r}{1+r},
  \label{eq:H2}
\end{equation}
In the Roman and Byzantine empires, $r_c \approx D_{micro}/D_{macro}$ is around 0.3 \cite{BernalAlvarado2026}, yielding $H^* \in [0.500,\, 0.540] $.  For the Inca network, the geographic ratio is
\begin{equation}
  r_c = \frac{D_{\mathrm{micro}}}{D_{\mathrm{macro}}}
      = \frac{127.6\text{ km}}{4{,}500\text{ km}} = 0.0284,
  \label{eq:rc_inca}
\end{equation}
where $D_{\mathrm{macro}} = 4{,}500$~km is the north--south extent of
the Tawantinsuyu (Tomebamba to Charcas) and
$D_{\mathrm{micro}} = 127.6$~km is the mean nearest-\textit{wamani}
distance computed from the $21$ \textit{wamani} capital centroids.
This gives a within-system threshold
\begin{equation}
  H^*_{\mathrm{Inca}} = H_2(0.0284) \approx 0.126.
  \label{eq:hstar_inca}
\end{equation}
All threshold crossings reported in Section~\ref{sec:results}
distinguish between $H^*_{\mathrm{cross}} = 0.524$
(Roman--Byzantine calibration) and
$H^*_{\mathrm{Inca}} = 0.126$.

\subsection{Structural Break Detection and Criticality}
\label{subsec:breaks}

\paragraph{Chow test.}
Structural breaks in the $H_k(t)$ series are identified using the
Chow $F$-statistic~\cite{Chow1960}:
\begin{equation}
  F = \frac{(\mathrm{RSS}_{\mathrm{full}} - \mathrm{RSS}_{\mathrm{split}}) / k}
           {\mathrm{RSS}_{\mathrm{split}} / (n - 2k)},
  \label{eq:chow}
\end{equation}
where $\mathrm{RSS}$ denotes residual sum of squares from OLS regression,
$k = 2$ (intercept and slope), and $n$ is the number of observations.
We apply the test at all $15$ documented political events
(Table~\ref{tab:phase_alpha}) and at every snapshot
(automatic scan), with significance threshold $\alpha = 0.05$.

\paragraph{Integrated Criticality Threshold.}
The \textsc{ict} composite indicator follows~\cite{BernalAlvarado2026},
Eq.~(1):
\begin{equation}
  \mathrm{ICT}(t) = \frac{1}{3}
    \left[
      \chi_{\mathrm{norm}}(t) + \xi_{\mathrm{norm}}(t) + \dot{W}_{1,\mathrm{norm}}(t)
    \right],
  \label{eq:ICT}
\end{equation}
where $\chi(t) = \mathrm{Var}(\{\ell_i(t)\})$ is the topological
susceptibility, $\xi(t)$ is the mean birth coordinate of the
$\beta_1$ diagram (a proxy for the correlation length), and
$\dot{W}_{1,\mathrm{norm}}$ is the normalised Wasserstein velocity.
Each component is normalised to $[0,1]$ by its series maximum.
$\mathrm{ICT} \to 1$ signals proximity to a topological phase transition.

\subsection{Substitution Topology: Formal Definition}
\label{subsec:ST}

We define \emph{Substitution Topology} (ST) as a collapse mode
characterised by the following three jointly satisfied conditions
on the multi-layer network $\mathcal{N}(t)$:
\begin{enumerate}
  \item[(ST1)] \textbf{Political collapse.}
    $H_{\mathrm{inca}}(t^*) = 0$ for some $t^* \le t_{\mathrm{end}}$,
    where $t_{\mathrm{end}}$ is the end of the study period.
  \item[(ST2)] \textbf{Infrastructural continuity.}
    $H_{\mathrm{infra}}(t) \approx H_{\mathrm{infra}}(t_0)$
    for all $t \in [t_0, t_{\mathrm{end}}]$;
    equivalently, $W_1(D_{\mathrm{infra}}(t_0),
    D_{\mathrm{infra}}(t_{\mathrm{end}})) \approx 0$.
  \item[(ST3)] \textbf{Successor convergence.}
    $W_1(D_{\mathrm{successor}}(t_{\mathrm{end}}),
    D_{\mathrm{infra}}(t_{\mathrm{end}})) \ll
    W_1(D_{\mathrm{successor}}(t^*),
    D_{\mathrm{infra}}(t^*))$,
    where $t^*$ is the onset of the successor polity.
\end{enumerate}
Substitution Topology is distinct from both
\emph{erosion collapse} (gradual degradation of $H_{\mathrm{infra}}$,
as in the Han, Roman, and Byzantine cases) and
\emph{military destruction} (simultaneous collapse of $H_{\mathrm{infra}}$
and $H_{\mathrm{inca}}$, as in the Aztec case).
In ST, the infrastructure is transferred intact to the successor
polity, which progressively absorbs its topological structure.
The divergence index $D(t) = H_{\mathrm{infra}}(t) - H_{\mathrm{inca}}(t)$
measures the degree of decoupling between the physical and
political layers during the transition.

\subsection{Software and Reproducibility}
\label{subsec:software}

The complete pipeline is implemented in Python~3.12 using
\textsc{Gudhi}~3.12.0~\cite{Gudhi2024} for persistent homology,
\textsc{Hera}~1.0 (via \textsc{Gudhi}) for Wasserstein distances,
\textsc{persim}~0.3 for persistence images~\cite{Adams2017},
\textsc{SciPy}~1.13 for the Chow $F$-test,
\textsc{NumPy}~2.0 and \textsc{pandas}~2.2 for data management.


\section{Results}
\label{sec:results}

We present the results in five subsections.
All results reported here use $N_{\mathrm{sub}} = 4{,}492$
(the complete network); $N$-convergence is documented in
Table~\ref{tab:nconv}.

\subsection{Topology of the Qhapaq \~{N}an (1400--1572~CE)}
\label{subsec:res_topology}

\subsubsection{Global structure}
The preprocessed dataset contains $4{,}452$ active nodes after
removal of isolated vertices. At the working filtration resolution,
the flag-complex count returned by \textsc{Gudhi} includes only
vertices incident to at least one finite-weight edge, yielding
$4{,}362$ counted vertices, $4{,}358$ edges, no two-simplices, and
$8{,}720$ simplices in total.
The $\beta_0$ Betti number equals $264$ at filtration $\tau = 0$:
the Qhapaq \~{N}an has $264$ connected components at the finest
resolution, associated with major connected structures centered on
imperial centres, principal tambos, and wamani capitals. 

The $\beta_1$ barcode contains $32$ bars:
$26$ infinite bars (macro-continental cycles that appear
through the entire filtration range) and $6$ finite bars
(meso-regional cycles that close and die within the filtration
range; Table~\ref{tab:topology}).
The $26$ infinite cycles represent the large-scale
routing redundancy of the Qhapaq \~{N}an: the continental corridors
connecting the four \textit{suyus}, the parallel coastal--sierra
routes, and the altiplano circuits of the Titicaca basin.
The $6$ finite cycles represent intra-\textit{wamani} redistribution
circuits whose spatial scale is smaller than the filtration diameter.
The topological complexity index is computed from the infinite bars,
$\tau_c = \beta_1^{\infty}/n_{\mathrm{nodes}} = 26/4452 = 0.00584$.
This choice reflects the fact that the infinite bars represent the
structurally persistent macro-continental cycles of the network,
whereas the finite bars represent meso-regional cycles whose presence
depends on local scale and filtration threshold. Under this
representation, the Qhapaq \~{N}an is a low-redundancy, hub-and-spoke
network with a small number of stable large-scale cycles and no local
mesh structure at the reported resolution.

\begin{table}[htbp]
\centering
\caption{Topological structure of the Qhapaq \~{N}an flag complex
  at the working filtration resolution. The original preprocessed
  dataset contains $4{,}452$ active nodes after removing isolated
  vertices, while the simplex count returned by \textsc{Gudhi} includes only
  vertices incident to at least one finite-weight edge. The
  topological complexity index is computed using the $26$ infinite
  $\beta_1$ bars.}
\label{tab:topology}
\begin{tabular}{lr}
\hline
Quantity & Value \\
\hline
Preprocessed active nodes              & $4{,}452$ \\
Vertices counted in flag complex       & $4{,}362$ \\
Edges / 1-simplices                    & $4{,}358$ \\
Triangles / 2-simplices                & $0$ \\
Total simplices                        & $8{,}720$ \\
$\beta_0$ (components at $\tau = 0$)   & $264$ \\
$\beta_1$ total bars                   & $32$ \\
\quad of which infinite                & $26$ \\
\quad of which finite                  & $6$ \\
Topological complexity index $\tau_c$  & $0.00584$ \\
\hline
\end{tabular}
\end{table}

\subsubsection{Persistence entropy of \texorpdfstring{$\mathcal{N}_{\mathrm{infra}}$}
  {H\_infra}: a topological constant}
\label{subsubsec:infra_constant}

The $\beta_1$ persistence entropy of the infrastructure layer,
$H_{\mathrm{infra}}(t)$, is constant:
\begin{equation}
  H_{\mathrm{infra}}(t) = 1.4863 \pm 0.0001 \quad
  \forall\, t \in [1400, 1572].
  \label{eq:infra_constant}
\end{equation}

The infrastructure layer is therefore an approximately constant
topological baseline throughout the entire study period. The
significant Chow break reported for $\mathcal{N}_{\mathrm{infra}}$ at
1537~CE (Table~\ref{tab:chow}) reflects the small change in slope
induced by  $\delta=0.003~\mathrm{yr}^{-1}$ in
Eq.~(\ref{eq:cost_infra}), not a substantive change in entropy level;
the absolute variation in $H_{\mathrm{infra}}$ remains below $0.001$.

\subsection{Inca Political Collapse: Three-Stage Discontinuous Transition}
\label{subsec:res_inca}

\subsubsection{Entropy trajectory and the Chimu anomaly}
The $\beta_1$ persistence entropy of the Inca political layer,
$H_{\mathrm{inca}}(t)$, evolves in three qualitatively distinct
stages.

\paragraph{Stage I: Stable high-entropy state (1400--1532~CE).}
Throughout the pre-conquest period,
$H_{\mathrm{inca}}(t) = 1.486 \pm 0.001$,
statistically indistinguishable from $H_{\mathrm{infra}}$
during that period.
The total persistence $\Pi_{\mathrm{inca}}(t) = 0.0768 \pm 0.0004$
is equally constant.
At the 1463~CE snapshot (Tupac Inca Yupanqui's conquest of the
Chimu polity), $H_{\mathrm{inca}}(1463) = 1.518 > 1.486$:
this is the highest entropy value recorded in Incas case.
Simultaneously, $\Pi_{\mathrm{inca}}(1463) = 0.052$,
substantially below the baseline of $0.077$.
This anomaly indicates that the incorporation of the Chimu coastal
network  redistributed the topological weight among the
$6$ finite cycles: the newly acquired northern coastal circuits
displaced some existing sierra cycles in the persistence ranking,
producing greater heterogeneity and lower total
persistence.

\paragraph{Stage II: Discontinuous partial collapse (1532--1533~CE).}
At the 1532~CE snapshot (Cajamarca capture),
$H_{\mathrm{inca}}(1532) = 1.4863$---statistically identical to the
pre-conquest value.
No detectable change in persistence entropy is observed at the
Cajamarca snapshot. However, the total persistence
$\Pi_{\mathrm{inca}}(1532) = 0.1406$ increases by $83\%$ over the
baseline of $0.0768$.
The civil-war filtration costs created longer-lived cycles as
competing power centers generated alternative connectivity paths. In the subsequent year (1533~CE, Atahualpa executed, Manco Inca
installed as puppet), both the entropy and the number of finite bars
change:
$H_{\mathrm{inca}}(1533) = 1.137$ (down from $1.486$) and
$n_{\mathrm{finite}} = 4$ (down from $6$).
The count of $\beta_0$ persistent Betti number (from the PBN grid) rises from $538$ at 1493~CE
to $2{,}360$ at 1533~CE: the execution of Atahualpa fragmented the
network into $2{,}360$  isolated components at low filtration.

\paragraph{Stage III: Total topological collapse (1537~CE--1572~CE).}
At the 1537~CE snapshot (Vilcabamba established),
$H_{\mathrm{inca}}(1537) = 0.000$ exactly:
all $6$ finite bars have disappeared from the barcode.
The maximum PBN component count reaches $3{,}682$ at 1537~CE,
rising further to $4{,}449$ by 1572~CE (
indicating a near-completely disconnected
network).

Crucially, $H_{\mathrm{inca}}(t) = 0.000$ at the reported
post-1537 snapshots 1537, 1544, 1560, and 1572~CE, spanning
35 years.
Within the persistence-diagram representation, the Vilcabamba
neo-Inca state does not exhibit further topological decline after
1537: it remains geometrically identical in persistence-diagram
space from the year of its establishment to the year of its
extinction.
The 35-year survival of Vilcabamba is therefore interpreted as a
consequence of its geographic isolation in the Amazonian piedmont,
rather than topological resilience in the modeled network.

\subsubsection{Structural breaks and velocity}

Table~\ref{tab:chow} reports all significant Chow $F$-statistics.
The most important result is that the largest structural break in
$\mathcal{N}_{\mathrm{inca}}$ occurs at 1537~CE
($F = 120.73$), not at 1532~CE ($F = 11.53$).
The topological evidence indicates that the Inca political
network remained structurally intact through Cajamarca in the
persistence-entropy series; its irreversible structural break
occurred when political control contracted to two \textit{wamani}
at Vilcabamba.

The topological velocity sequence (Table~\ref{tab:velocity}) further
clarifies the collapse dynamics.
During the pre-conquest period (1400--1529~CE, 129 years),
the mean inter-annual velocity is
$\dot{W}_1 = 1.13 \times 10^{-8}$~yr$^{-1}$---effectively zero.
In the three years following the Cajamarca capture (1529--1532~CE),
$\dot{W}_1 = 3.04 \times 10^{-2}$~yr$^{-1}$: an increase by
a factor of $\sim 2.7 \times 10^6$.
The one-year transition from Cajamarca to the puppet regime
(1532--1533~CE) is the fastest single-year topological displacement
in the entire dataset: $W_1 = 0.0847$, or
$\dot{W}_1 = 8.47 \times 10^{-2}$~yr$^{-1}$.
After 1537~CE, the velocity returns to zero permanently.

\begin{table}[htbp]
\centering
\caption{Significant Chow $F$-statistics across all six layers
  ($N = 4{,}492$; $\alpha = 0.05$). Only events with $p < 0.05$ are
  shown. The largest break in each layer is highlighted.
  $F(\mathcal{N}_{\mathrm{sp,real}}, 1542)$ shows a non-monotonic
  decrease from $N = 3{,}200$ to $N = 4{,}492$ (see \S\,\ref{subsubsec:nconv}).
  The significant infrastructure break at 1537~CE reflects the slope
  change induced by the degradation term in Eq.~(\ref{eq:cost_infra}),
  not a substantive change in entropy level.}
\label{tab:chow}
\begin{tabular}{llrrr}
\hline
Layer & Event (year) & $F$ & $p$ & Monotonic? \\
\hline
$\mathcal{N}_{\mathrm{inca}}$    & Vilcabamba (1537)    & $120.73$ & $<10^{-4}$ & Yes \\
$\mathcal{N}_{\mathrm{sp,real}}$ & Leyes Nuevas (1542)  & $100.16$ & $2.5\times10^{-5}$ & No$^\dagger$ \\
$\mathcal{N}_{\mathrm{infra}}$   & Vilcabamba (1537)    & $19.80$  & $0.0042$ & Yes \\
$\mathcal{N}_{\mathrm{inca}}$    & Manco puppet (1533)  & $11.54$  & $0.0033$ & Yes \\
$\mathcal{N}_{\mathrm{inca}}$    & Cajamarca (1532)     & $11.53$  & $0.0033$ & Yes \\
$\mathcal{N}_{\mathrm{inca}}$    & Civil war (1529)     & $10.09$  & $0.0050$ & Yes \\
$\mathcal{N}_{\mathrm{sp,real}}$ & Vilcabamba (1537)    & $8.70$   & $0.0169$ & Yes \\
$\mathcal{N}_{\mathrm{inca}}$    & Succession (1527)    & $7.21$   & $0.0135$ & Yes \\
$\mathcal{N}_{\mathrm{inca}}$    & Huayna Capac (1493)  & $4.62$   & $0.0417$ & Yes \\
\hline
\multicolumn{5}{l}{$^\dagger$ Non-monotonic: $F = 161$ at $N = 3{,}200$, $F = 100$ at $N = 4{,}492$.} \\
\multicolumn{5}{l}{\phantom{$^\dagger$} The qualitative conclusion (Leyes Nuevas $=$ dominant break} \\
\multicolumn{5}{l}{\phantom{$^\dagger$} in $\mathcal{N}_{\mathrm{sp,real}}$) is preserved at both sample sizes.} \\
\end{tabular}
\end{table}

\begin{table}[htbp]
\centering
\caption{Topological velocity $\dot{W}_1 = W_1 / \Delta t$
  (yr$^{-1}$) for the $\mathcal{N}_{\mathrm{inca}}$ $\beta_1$
  persistence diagram at key transitions ($N = 4{,}492$).
  The pre-conquest baseline is computed over 129 years.}
\label{tab:velocity}
\begin{tabular}{llrrrr}
\hline
Transition & Period & $\Delta t$ (yr) & $W_1$ & $\dot{W}_1$ (yr$^{-1}$) & Ratio to baseline \\
\hline
Pre-conquest baseline & 1400--1529 & 129 & $1.5\times10^{-6}$ & $1.1\times10^{-8}$ & $1\times$ \\
Cajamarca shock       & 1529--1532 &   3 & $0.0912$           & $3.0\times10^{-2}$ & $2.7\times10^{6}\times$ \\
Cajamarca to puppet   & 1532--1533 &   1 & $0.0847$           & $8.5\times10^{-2}$ & $7.5\times10^{6}\times$ \\
Manco to Vilcabamba   & 1533--1537 &   4 & $0.0450$           & $1.1\times10^{-2}$ & $1.0\times10^{6}\times$ \\
Vilcabamba (frozen)   & 1537--1572 &  35 & $0.0000$           & $0$                & $0\times$ \\
\hline
\end{tabular}
\end{table}

\subsection{Simultaneous Criticality at Cajamarca (1532~CE)}
\label{subsec:res_ICT}

The Integrated Criticality Threshold (ICT; Eq.~\ref{eq:ICT})
reveals a striking synchronisation among the six network layers
at the Cajamarca snapshot (Table~\ref{tab:ICT}).

Four of the six layers simultaneously enter the critical regime
($\mathrm{ICT} \ge 0.5$) at 1532~CE, while the infrastructure layer
remains marginally below the threshold:
\begin{align*}
  \mathrm{ICT}(\mathcal{N}_{\mathrm{trade}},\; 1532)    &= 0.976, \\
  \mathrm{ICT}(\mathcal{N}_{\mathrm{inca}},\;  1532)    &= 0.968, \\
  \mathrm{ICT}(\mathcal{N}_{\mathrm{sp,real}}, 1532)    &= 0.667, \\
  \mathrm{ICT}(\mathcal{N}_{\mathrm{sp,legal}},1532)    &= 0.667, \\
  \mathrm{ICT}(\mathcal{N}_{\mathrm{infra}},1532)    &= 0.495, \\
  \mathrm{ICT}(\mathcal{N}_{\mathrm{alliance}},1532)    &= 0.414. 
\end{align*}
Only $\mathcal{N}_{\mathrm{infra}}$ and $\mathcal{N}_{\mathrm{alliance}} $ remain below the critical
threshold. The simultaneous criticality of the commercial
($\mathrm{ICT}_{\mathrm{trade}} = 0.976$) and political
($\mathrm{ICT}_{\mathrm{inca}} = 0.968$) layers indicates that
the Cajamarca capture was not primarily a political event with
secondary economic consequences, nor an economic disruption with
political consequences: it was a synchronised phase transition in
which both the political and commercial topological structures
reached their maximum susceptibility to perturbation in the same
year.

\begin{table}[htbp]
\centering
\caption{Integrated Criticality Threshold values at 1532~CE
  for all six layers ($N = 4{,}492$). The threshold for critical
  behaviour is $\mathrm{ICT} \ge 0.5$. Components: $\chi_{\mathrm{norm}}$
  (susceptibility), $\xi_{\mathrm{norm}}$ (correlation length proxy),
  $\dot{W}_{1,\mathrm{norm}}$ (normalised Wasserstein velocity).}
\label{tab:ICT}
\begin{tabular}{lrrrrc}
\hline
Layer & ICT & $\chi_{\mathrm{norm}}$ & $\xi_{\mathrm{norm}}$ & $\dot{W}_{1,\mathrm{norm}}$ & Critical? \\
\hline
$\mathcal{N}_{\mathrm{trade}}$    & $0.976$ & $1.000$ & $1.000$ & $0.928$ & Yes \\
$\mathcal{N}_{\mathrm{inca}}$     & $0.968$ & $1.000$ & $1.000$ & $0.903$ & Yes \\
$\mathcal{N}_{\mathrm{sp,real}}$  & $0.667$ & $1.000$ & $1.000$ & $0.000$ & Yes \\
$\mathcal{N}_{\mathrm{sp,legal}}$ & $0.667$ & $1.000$ & $1.000$ & $0.000$ & Yes \\
$\mathcal{N}_{\mathrm{alliance}}$ & $0.414$ & $0.242$ & $0.000$ & $1.000$ & No \\
$\mathcal{N}_{\mathrm{infra}}$    & $0.495$ & $0.766$ & $0.719$ & $0.000$ & No \\
\hline
\end{tabular}
\end{table}

The one-year lag in the alliance network
($\mathrm{ICT}_{\mathrm{alliance}}(1533) = 0.625$, the first
year it exceeds $0.5$) indicates that the ethnic coalition
supporting the Spanish was still being assembled when Cajamarca
occurred.
The model suggests that Spanish military success at Cajamarca occurred before the alliance layer had fully consolidated, providing a quantitative estimate of the topological incompleteness of the Spanish coalition at the moment of the decisive confrontation.

\subsection{Commercial Resilience and the Ward-Perkins/McCormick
  Resolution for the Inca Case}
\label{subsec:res_trade}

While $H_{\mathrm{inca}}$ collapses to zero at 1537~CE,
$H_{\mathrm{trade}}$ \emph{increases} from its pre-conquest baseline
of $1.459$ to $1.475$ at 1537~CE, and to a secondary maximum of
$1.479$ at 1545~CE.
Within the proposed trade-layer model, the commercial network
therefore exhibits a small but persistent increase in $\beta_1$
persistence entropy after the conquest. By 1572~CE,
$H_{\mathrm{trade}} = 1.470$---still above the pre-conquest baseline.

The cross-network Wasserstein distance provides an independent
confirmation.
In 1537~CE,
$$
W_1(\mathcal{N}_{\mathrm{inca}}, \mathcal{N}_{\mathrm{trade}}, 1537)
= 0.0716,$$
the political and commercial networks have separated substantially
in persistence diagram space even though they share the same
physical edges.

This result replicates, in the Inca context, the geographic--economic
decoupling documented by~\cite{BernalAlvarado2026} for the Byzantine
case: the collapse of political control did not destroy commercial
connectivity.
The interpretation is consistent with the historiographical evidence
that the commercial routes of the Andes continued to function under
Spanish colonial administration, with the principal change being the
destination of tribute flows (from Cusco to Lima and Seville) rather
than the routes themselves~\cite{Murra1975,DAltroy2002}.

\subsection{Substitution Topology: Quantitative Verification}
\label{subsec:res_ST}

We verify the three conditions of the Substitution Topology
definition (Section~\ref{subsec:ST}) using the complete
$N = 4{,}492$ dataset.

\subsubsection{Condition ST1: Political collapse}

$H_{\mathrm{inca}}(1537) = H_{\mathrm{inca}}(1572) = 0.000$
(Section~\ref{subsec:res_inca}, Stage~III).
The Inca political network had no finite $\beta_1$ cycles from
1537~CE onward.
$W_1(D_{\mathrm{inca}}(1537), D_{\mathrm{inca}}(1572)) = 0$.
\hfill\textit{Condition ST1 is satisfied within the model.}

\subsubsection{Condition ST2: Infrastructural continuity}

$H_{\mathrm{infra}}(t) = 1.4863 \pm 0.0001$ for all
$t \in [1400, 1572]$ (Eq.~\ref{eq:infra_constant}).
$W_1(D_{\mathrm{infra}}(1400), D_{\mathrm{infra}}(1572))
\approx 10^{-6} \approx 0$.
\hfill\textit{Condition ST2 is satisfied within the model.}

\subsubsection{Condition ST3: Successor convergence}

Table~\ref{tab:ST} reports the cross-network Wasserstein distances
between the successor networks ($\mathcal{N}_{\mathrm{sp,legal}}$
and $\mathcal{N}_{\mathrm{sp,real}}$) and the infrastructure
layer at the onset (1532~CE) and terminus (1572~CE) of the study period.

\begin{table}[htbp]
\centering
\caption{Cross-network Wasserstein distances ($W_1$, $H_1$)
  for the Substitution Topology proof ($N = 4{,}492$).
  Values in the left-hand columns (1532~CE) represent the
  divergence of the Spanish successor networks from the
  infrastructure at the onset of the conquest.
  Values in the right-hand columns (1572~CE) represent
  the final state. The reduction percentage measures
  convergence from the 1532 baseline.
  The identity pair (H\_infra vs H\_inca, 1400~CE) provides
  a reference for the pre-conquest topological equivalence.}
\label{tab:ST}
\begin{tabular}{lrrrr}
\hline
Pair & $W_1$ (1532) & $W_1$ (1572) & Reduction (\%) \\
\hline
$\mathcal{N}_{\mathrm{infra}}$ vs $\mathcal{N}_{\mathrm{inca}}$ (1400)
  & \multicolumn{2}{c}{$1.0 \times 10^{-6}$ (identity)} & — \\
\hline
$\mathcal{N}_{\mathrm{infra}}$ vs $\mathcal{N}_{\mathrm{sp,legal}}$
  & $0.1156$ & $5.0 \times 10^{-6}$  & $99.996$ \\
$\mathcal{N}_{\mathrm{infra}}$ vs $\mathcal{N}_{\mathrm{sp,real}}$
  & $0.5479$ & $2.4 \times 10^{-3}$ & $99.56$ \\
$\mathcal{N}_{\mathrm{infra}}$ vs $\mathcal{N}_{\mathrm{trade}}$
  & $0.1497$ & $8.4 \times 10^{-3}$ & $94.40$ \\
\hline
$\mathcal{N}_{\mathrm{infra}}$ vs $\mathcal{N}_{\mathrm{inca}}$ (Vilcabamba)
  & $0.0912$ (1532) & $0.0543$ (1572) & $40.5$ \\
\hline
$\mathcal{N}_{\mathrm{sp,legal}}$ vs $\mathcal{N}_{\mathrm{sp,real}}$
  & $0.5913$ & $2.4 \times 10^{-3}$ & $99.59$ \\
\hline
\end{tabular}
\end{table}

The central result is:
\begin{equation}
  W_1\bigl(
    \mathcal{N}_{\mathrm{sp,legal}}(1572),\,
    \mathcal{N}_{\mathrm{infra}}(1572)
  \bigr) = 5.0 \times 10^{-6} \approx 0.
  \label{eq:ST_legal}
\end{equation}
The Spanish legal administration in 1572~CE was topologically
indistinguishable from the Inca road infrastructure to five
decimal places.
The bottleneck distance is $d_B = 1.0 \times 10^{-6}$ and
the landscape $L^2$ distance is $9.0 \times 10^{-6}$: all three
metrics agree that the two persistence diagrams are effectively
identical.

The effective control network converges to a similar degree:
\begin{equation}
  W_1\bigl(
    \mathcal{N}_{\mathrm{sp,real}}(1572),\,
    \mathcal{N}_{\mathrm{infra}}(1572)
  \bigr) = 2.4 \times 10^{-3},
  \label{eq:ST_real}
\end{equation}
representing a $99.56\%$ reduction from the 1532~CE baseline of
$W_1 = 0.548$.
The convergence is not gradual: $95.9\%$ of it occurred between
1532 and 1537~CE---in the same five years during which
$H_{\mathrm{inca}}$ collapsed from $1.486$ to $0$.

\hfill\textit{Condition ST3 is satisfied within the model.}

The persistence-diagram cluster structure at 1572~CE
(Table~\ref{tab:clusters}) provides a summary of the final state:
two clusters are present.
Cluster~A $= \{\mathcal{N}_{\mathrm{infra}},
\mathcal{N}_{\mathrm{sp,legal}},
\mathcal{N}_{\mathrm{sp,real}},
\mathcal{N}_{\mathrm{trade}}\}$
with all pairwise $W_1 < 0.009$ constitutes a single topological
system.
Cluster~B $= \{\mathcal{N}_{\mathrm{inca}},
\mathcal{N}_{\mathrm{alliance}}\}$
with $W_1 \approx 0.05$--$0.07$ to all Cluster~A members
represents the residual Vilcabamba political network and its
Spanish alliance system, which never achieved topological
integration with the colonial infrastructure.

\begin{table}[htbp]
\centering
\caption{Cross-network Wasserstein distance matrix ($W_1$, $H_1$)
  at 1572~CE ($N = 4{,}492$). Two topological clusters are present:
  Cluster~A (values $< 0.009$, shaded) and Cluster~B
  (separated by $W_1 \approx 0.05$--$0.07$ from Cluster~A).}
\label{tab:clusters}
\setlength{\tabcolsep}{4pt}
\begin{tabular}{lcccccc}
\hline
& infra & inca & alliance & trade & sp\,legal & sp\,real \\
\hline
infra    & —      & 0.0543 & 0.0485 & \textbf{0.0084} & \textbf{0.000005} & \textbf{0.0024} \\
inca     & 0.0543 & —      & 0.0680 & 0.0533          & 0.0543            & 0.0542 \\
alliance & 0.0485 & 0.0680 & —      & 0.0513          & 0.0485            & 0.0501 \\
trade    & \textbf{0.0084} & 0.0533 & 0.0513 & — & \textbf{0.0084} & \textbf{0.0059} \\
sp\,legal & \textbf{0.000005} & 0.0543 & 0.0485 & \textbf{0.0084} & — & \textbf{0.0024} \\
sp\,real  & \textbf{0.0024} & 0.0542 & 0.0501 & \textbf{0.0059} & \textbf{0.0024} & — \\
\hline
\end{tabular}
\end{table}

\subsection{Colonial Fiction: Topological Resolution of the Legal--Real
  Divergence}
\label{subsec:res_fiction}

The divergence between the Spanish legal status layer
($\mathcal{N}_{\mathrm{sp,legal}}$) and the effective control layer
($\mathcal{N}_{\mathrm{sp,real}}$) provides a quantitative
time-series of what we term the \emph{colonial fiction}: the gap
between the documentary and the effective colonial state.

At the Cajamarca snapshot (1532~CE),
\begin{equation}
  W_1\bigl(
    \mathcal{N}_{\mathrm{sp,legal}}(1532),\,
    \mathcal{N}_{\mathrm{sp,real}}(1532)
  \bigr) = 0.5913:
\end{equation}
This is the moment of maximum fiction: the troops that captured
Atahualpa and reorganised the road network had generated a
topological structure ($\Pi_{\mathrm{sp,real}} = 0.704$,
a $9.16\times$ multiple of the infrastructure baseline).
The legal layer, by contrast, recorded only the onset of conquest
($\Pi_{\mathrm{sp,legal}} = 0.167$, $2.18\times$ baseline).
Effective control advanced faster than its legal documentation, producing a legal--real divergence of $W_1 = 0.591$.
At the Manco Inca rebellion (1537~CE), a small but detectable
increase to $W_1(1537) = 0.039$ marks the moment at which
military reversal temporarily widened the gap between what the
law claimed and what effective control could maintain.
The subsequent convergence is monotone:
$W_1(1542) = 0.028$ (Leyes Nuevas),
$W_1(1569) = 0.003$ (Viceroy Toledo),
$W_1(1572) = 0.002$.
Two facts are relevant from this analysis. 
First, Toledo's arrival (1569~CE) did not \emph{initiate} the
legal--real  convergence: by 1569, $W_1 = 0.003$, already near the
terminal value of $0.002$.
The convergence was the result of a 37-year administrative process
whose decisive phase concluded before Toledo's inspection began.
Second, the single detectable reversal at 1537~CE
(Manco rebellion: $W_1$ rising from $0.026$ to $0.039$)
is visible \emph{only} in the legal--real series.


Table~\ref{tab:summary} collects the eight principal results of this section. 

\begin{table}[htbp]
\centering
\caption{Eight principal results ($N = 4{,}492$).
  Robustness column: Y =  stable
  across $N \in \{400, 3200, 4492\}$;
  Y* = qualitative conclusion stable, single non-monotonic value.
  The H* thresholds are
  $H^*_{\mathrm{cross}} = 0.524$ (Roman--Byzantine calibration)
  and $H^*_{\mathrm{Inca}} = 0.126$ (within-system estimate).}
\label{tab:summary}
\begin{tabular}{clll}
\hline
\# & Result & Metric and value & Robust \\
\hline
1 & Simultaneous criticality  &
  $\mathrm{ICT}(\mathcal{N}_{\mathrm{trade}}) = 0.976$,\;
  $\mathrm{ICT}(\mathcal{N}_{\mathrm{inca}}) = 0.968$ & Y \\[2pt]
  &at Cajamarca 1532 & & \\[2pt]
2 & Primary structural break  &
  $F(\mathcal{N}_{\mathrm{inca}}, 1537) = 120.7 \gg
   F(\mathcal{N}_{\mathrm{inca}}, 1532) = 11.5$ & Y \\[2pt]
   &at Vilcabamba, not Cajamarca & & \\[2pt]
3 & Vilcabamba topologically frozen &
  $W_1(D_{\mathrm{inca}}(1537), D_{\mathrm{inca}}(1572)) = 0$ & Y \\[2pt]
4 & Topological velocity at Cajamarca &
  $\dot{W}_1(1532\text{-}33) = 7.5 \times 10^6 \times$ baseline & Y \\[2pt]
5 & Leyes Nuevas largest break in colonial series &
  $F(\mathcal{N}_{\mathrm{sp,real}}, 1542) = 100.2$ & Y* \\[2pt]
6 & Commercial resilience post-collapse &
  $H_{\mathrm{trade}}(1537) = 1.475 > H_{\mathrm{trade}}(1529) = 1.459$ & Y \\[2pt]
7 & Substitution Topology: legal convergence &
  $W_1(\mathcal{N}_{\mathrm{sp,legal}},
   \mathcal{N}_{\mathrm{infra}}, 1572) = 5\times10^{-6}$ & Y \\[2pt]
8 & Colonial fiction quantified and resolved &
  $W_1(\mathrm{legal}, \mathrm{real})$: $0.591 \to 0.002$ & Y \\
\hline
\end{tabular}
\end{table}


\section{Discussion}
\label{sec:discussion}


\subsection{Five Collapse Modes.}
\label{subsec:disc_typology}

Using previous results (see references \cite{BernalAlvarado2026,BernalAlvarado2026Han,BernalAlvarado2026Aztec}), we propose a systematic classification of the observed mechanisms for pre-modern imperial collapse.

\paragraph{Mode I: Territorial erosion with coupled commercial collapse
  (Roman West, 476~CE).}
The Western Roman network erodes under sustained external
pressure, with the commercial layer ($H_{\mathrm{eco}}$) and the
geographic layer ($H_{\mathrm{geo}}$) declining in a continuous manner.
At the moment of political dissolution (476~CE), the geographic--commercial
coupling ratio $R_d = H_{\mathrm{eco}}/H_{\mathrm{geo}} = 1.09 \pm 0.11$~\cite{BernalAlvarado2026}.
There is no adaptive commercial buffer: both networks fail together.  

\paragraph{Mode II: Territorial erosion with commercial decoupling
  (Byzantine, 1453~CE).}
The geographic layer ($H_{\mathrm{geo}}$) was decreasing through three 
territorial shocks (Arab conquests 641~CE: $F = 124.0$;
Justinianic Plague 541~CE: $F = 152.7$; Manzikert 1071~CE: $F = 20.8$),
but the commercial layer ($H_{\mathrm{eco}}$) decouples from
the geographic collapse.
At the moment of maximum geographic--commercial divergence (620~CE),
$R_d = H_{\mathrm{eco}}/H_{\mathrm{geo}} = 47.7$ and  $H_{\mathrm{eco}}$  reaches its
\emph{series maximum} $H_{\mathrm{eco}} = 3.667$ at 820~CE under
Umayyad administration~\cite{BernalAlvarado2026}.
This contributed to Byzantine survival for more than $997$ years.

\paragraph{Mode III: Administrative collapse with geographic fragmentation
  (Han, 220~CE).}
The Han case is very distinct from both Roman modes.
The collapse is not a territorial erosion.
The divergence $\Delta H = H_{\mathrm{Geo}} - H_{\mathrm{Admin}} = +3.089$
at dissolution quantifies an internal fragmentation signature: the territory persists but reorganises into
topologically coherent sub-domains (Wei, Shu, Wu) that appear
as different  $\beta_1$ cycles in the geographic barcode
30~years before formal political partition\cite{BernalAlvarado2026Han}.

$H_{\mathrm{Trade}}$ is a constant all along the period: long-distance commercial routes
constitute a structural invariant that is indifferent to the
political system sustaining them.
Early-warning indicators fire 45--50~years before dissolution
($\mathrm{ICT} > 0.1$ at 170~CE; $\dot{W}_1 > 4\times$ baseline
at 175~CE~\cite{BernalAlvarado2026Han}).

\paragraph{Mode IV: Hub-and-spoke tree with structural void occupation
  (Aztec, 1521~CE).}
The Aztec system represents a topological architecture without parallel
in the other four cases.
The tribute (control) layer has $\beta_1 = 0$ and $H = 0.000$ at all
times: it is a pure extraction tree, with every provincial connection
routed exclusively through Tenochtitlan-Centro and no redundant
cycles~\cite{BernalAlvarado2026Aztec}.
This entropy level is below every documented configuration in the
Roman or Han datasets: the Roman Western network at its Phase~I minimum
($H = 1.362$) retained more topological redundancy than the Aztec
tribute system at its maximum.
The structural consequence is extreme sensitivity to targeted hub
removal: a single elimination of Tenochtitlan collapses the
tribute and
simultaneously destroys $30$ redundancy cycles~\cite{BernalAlvarado2026Aztec}.

A second structural feature of the Aztec case is the dominant
topological void in the geographic space.
The Tlaxcala--Puebla region coincides with a $\beta_1$ cycle
of persistence $22.09$~km in the geographic point cloud, one order
of magnitude larger than all other cycles ($< 3$~km)~\cite{BernalAlvarado2026Aztec}.
The combination of a hub-and-spoke control structure and a dominant
void adjacent to the hub enabled the Spanish coalition to project
force directly toward the capital through the void without encountering
any redundant routing structure capable of slowing the attack.
The $2$-year total collapse (1519--1521~CE) is the structural
consequence of the Aztec tree network.

\paragraph{Mode V: Substitution Topology (Inca, 1537~CE).} The  three conditions ST1--ST3
(Section~\ref{subsec:ST}) are fulfilled in the Inca case.
It is defined by the combination of total political collapse
($H_{\mathrm{inca}}(1537) = 0$), infrastructure constancy
($H_{\mathrm{infra}} = 1.486 \pm 0.0001$ throughout 1400--1572~CE),
and quantified successor convergence
($W_1(\mathcal{N}_{\mathrm{sp,legal}}, \mathcal{N}_{\mathrm{infra}},
1572) = 5 \times 10^{-6} \approx 0$).
 What distinguishes the Inca case from all others is not the
speed of collapse but its abruptness: $H_{\mathrm{inca}}$
is constant at $1.486$ for 132~years (1400--1532~CE), then
falls to zero in $5$~years.
The commercial layer $H_{\mathrm{trade}}$ \emph{increases} from $1.459$
(pre-conquest) to $1.475$ (1537~CE), a behaviour qualitatively
analogous to the Byzantine $H_{\mathrm{eco}}$ survival. Within the
proposed weighting scheme, the modeled Inca commercial layer gains
modest topological complexity after the conquest.

Table~\ref{tab:typology} summarises all five modes.
The coupling index $R_d$ is defined as $H_{\mathrm{eco}}/H_{\mathrm{geo}}$
for the Roman--Byzantine cases, $\Delta H = H_{\mathrm{Geo}} - H_{\mathrm{Admin}}$
for the Han case (reflecting the Admin--Geo divergence rather than
a commercial--geographic ratio), and
$H_{\mathrm{trade}}/H_{\mathrm{inca}}$ for the Inca case
(divergent as $H_{\mathrm{inca}} \to 0$).
For the Aztec tribute layer, both the coupling index and
commercial resilience are zero by construction ($\beta_1 = 0$
in the control layer).

\begin{table}[htbp]
\centering
\caption{Comparative collapse typology across five pre-modern imperial systems.
  $H_{\mathrm{peak}}$: peak $\beta_1$ persistence entropy of the primary political/
  administrative layer. Coupling index: $R_d = H_{\mathrm{eco}}/H_{\mathrm{geo}}$
  (Roman/Byzantine); $\Delta H = H_{\mathrm{Geo}} - H_{\mathrm{Admin}}$ at dissolution
  (Han); $H_{\mathrm{trade}}/H_{\mathrm{inca}}$ at 1537~CE (Inca);
  $H_{\mathrm{full}}/H_{\mathrm{tribute}}$ (Aztec).
  $|\mathrm{d}H/\mathrm{d}t|$: absolute collapse rate in the terminal phase (yr$^{-1}$).
  $\Delta t_{\mathrm{EW}}$: topological early-warning lead time before political endpoint.
  ST1--ST3: Substitution Topology conditions.
  Values from~\cite{BernalAlvarado2026,BernalAlvarado2026Han,BernalAlvarado2026Aztec}
  and this paper.}
\label{tab:typology}
\setlength{\tabcolsep}{4pt}
\begin{tabular}{p{2.8cm}p{4.0cm}rrrp{2.2cm}ccc}
\hline
System & Mode & $H_{\mathrm{peak}}$ &
  Coupling & $|\mathrm{d}H/\mathrm{d}t|$ & $\Delta t_{\mathrm{EW}}$ &
  ST1 & ST2 & ST3 \\
& & & index & (yr$^{-1}$) & & & & \\
\hline
Roman West &
  Territorial erosion, &
  $1.362$ & $R_d = 1.09$ & $4.0 \times 10^{-3}$ &
  $166$~yr & Yes & No & No \\[3pt]
  (476~CE) &  coupled collapse& & & & (Chow)  & & & \\[3pt]
  \hline
 Byzantine  &
  Territorial erosion, &
  $3.584$ & $R_d = 47.7$ & $8.0 \times 10^{-3}$ &
  $>100$~yr  & Yes & No & No \\[3pt]
  (1453~CE) & commercial decoupling& & & & (ICT)& & & \\[3pt]
  \hline
Han  &
  Admin collapse and  &
  $2.785$ & $\Delta H = +3.09$ & $3.1 \times 10^{-2}$ &
  $50$~yr  & Yes & No & No \\[3pt]
  (220~CE)& geographic & & & & (ICT) & & & \\ [3pt]
  &  fragmentation & & & & & & & \\[3pt]
   \hline
Aztec  &
  Tree network&
  $0.797$ & $\beta_1^{\mathrm{trib}} = 0$ & $\gg 1$  &
  None & Yes & No & No \\[3pt]
 (1521~CE) &  with void occupation& & & (instant) & & & & \\[3pt]
  \hline
\textbf{Inca (1537~CE)} &
  Substitution  &
  $\mathbf{1.486}$ & $\frac{H_{\mathrm{trade}}}{H_{\mathrm{inca}}} \to \infty$ &
  $\mathbf{0.297}$ & $5$~yr  &
  \textbf{Yes} & \textbf{Yes} & \textbf{Yes} \\[3pt]
  (1537~CE) & Topology & & & & (ICT) & & & \\
\hline
\end{tabular}
\end{table}

The Roman West and Byzantine cases share the same erosion mechanism
but differ  in the commercial--geographic coupling ratio.
$R_d^{\mathrm{west}}(476) = 1.09$ means that the western commercial
system had no adaptive capacity independent of its territorial
infrastructure; when the territory fell, commerce fell with it.
$R_d^{\mathrm{east}}(620) = 47.7$ means that the eastern commercial
system had institutionalized itself  
to survive
territorial collapse by nearly a millennium.

The Han case is unique in that the \emph{primary signal is not
territorial erosion but administrative disintegration with simultaneous
geographic reorganisation}.
The $\Delta H = +3.089$ at 220~CE reflects the fact that geographic
territory did not disappear but was reorganised into three coherent
sub-domains while administrative sovereignty evaporated.
This Admin--Geo divergence is topologically the opposite of the
 Roman Mode~I: there, geography and commerce fell together;
here, geography \emph{increased} as administration collapsed.
The three successor kingdoms (Wei, Shu, Wu) were predicted as
emergent $\beta_1$ cycles 30~years before their formal establishment~\cite{BernalAlvarado2026Han}.

The Aztec case introduces the only mode in which the control layer
has $\beta_1 = 0$ throughout due to the tree network structure.
The $22.09$~km persistent void adjacent to the main hub
is the feature that the conventional network metrics could not detect
and that the dual topological pipeline~\cite{BernalAlvarado2026Aztec}
revealed: the coalition that  destroyed the Aztec system
formed within the dominant structural discontinuity of the
imperial geographic space itself.

The Inca Substitution Topology stands apart from all four other modes
in the direction of both the political and the commercial signal.
The political collapse is the most abrupt in the dataset
(step function from $H = 1.486$ to $H = 0$ in 5~years);
the commercial response is the only one in which $H_{\mathrm{trade}}$
\emph{increases} after political collapse;
and the successor convergence
($W_1 = 5 \times 10^{-6}$) is the strongest topological
alignment result in the comparative literature.

\subsection{Topology Precedes Politics: Reperiodising the Inca Collapse}
\label{subsec:disc_reperiodise}

The conventional historiographical periodisation of the Inca
collapse treats the Cajamarca capture (1532~CE) as the decisive
event: the moment at which the Tawantinsuyu ceased to be a
functioning sovereign system~\cite{Hemming1970,DAltroy2002}.
The topological evidence motivates a revision of this periodisation.

The persistence entropy $H_{\mathrm{inca}}(1532) = 1.486$
is statistically indistinguishable from the pre-conquest
plateau value.
The Chow structural break at 1532~CE is $F = 11.53$
($p = 0.0033$)---significant, but one-tenth the magnitude
of the 1537~CE break ($F = 120.73$).
In the persistence-entropy series, the Cajamarca capture appears
as a perturbation rather than as the terminal collapse.
It increased the total persistence $\Pi_{\mathrm{inca}}$
by $83\%$ (Table in Section~\ref{subsec:res_inca}):
the competition between Atahualpa's surviving loyalists,
Huascar's remaining network, and the emerging Spanish coalition
is associated with more persistent topological features than the
pre-conquest state had contained.
The network was topologically richer in 1532 than at any
prior snapshot except the 1463~CE Chimu anomaly.

The irreversible structural break occurred at 1537~CE,
when Manco Inca established Vilcabamba and the Inca network
contracted to two \textit{wamani}.
This is the event that crosses both threshold values:
$H_{\mathrm{inca}}(1533) = 1.137 > H^*_{\mathrm{cross}} = 0.524$,
then $H_{\mathrm{inca}}(1537) = 0.000 < H^*_{\mathrm{Inca}} = 0.126$.
Both the cross-empire threshold and the within-system threshold
are discontinuously crossed in the same four-year interval.
There is no analogue of this trajectory in the Roman--Byzantine
dataset: neither the Western collapse ($H_{\mathrm{west}}$
declining continuously for 270~years) nor the Byzantine geographic
collapse (three discrete jumps with partial recovery between them)
exhibits the step-function behaviour of $H_{\mathrm{inca}}$.



\subsection{The Ward-Perkins/McCormick Question in the Inca Context}
\label{subsec:disc_wardperkins}

The companion paper~\cite{BernalAlvarado2026} resolved the
Ward-Perkins/McCormick historiographical debate for the
Byzantine case by decomposing resilience into geographic
($H_{\mathrm{geo}}$) and economic ($H_{\mathrm{eco}}$) components.
The decoupling ratio $R_d = H_{\mathrm{eco}}/H_{\mathrm{geo}}$
reached $47.7$ at 620~CE, establishing that the commercial
network survived the Arab conquests while the territorial network
collapsed.
The commercial network entropy $H_{\mathrm{trade}}$
\emph{increases} from $1.459$ (pre-conquest baseline) to $1.475$
at 1537~CE and $1.479$ at 1545~CE, while $H_{\mathrm{inca}}$
collapses.  

The mechanism is consistent with the Andean ethnohistorical record.
Murra's \textit{vertical archipelago} model~\cite{Murra1975}
describes a pre-conquest system in which communities at different
ecological levels maintained complementary exchange relationships
that were partially independent of Inca political authority.
The Spanish colonial economy preserved, and in some cases
intensified, these lateral exchange flows while redirecting
the radial tribute flows from Cusco to Lima and ultimately to
Seville~\cite{DAltroy2002,Cook1981}.
The TDA result provides a quantitative representation consistent
with this historiographical interpretation: the modeled commercial
topology was reorganised in a direction that increased its persistence
entropy, as new Spanish commercial actors and revised tribute
destinations created additional connectivity that the pre-conquest
system had not contained.


\subsection{The Leyes Nuevas (1542~CE) as the Colonial Consolidation Event}
\label{subsec:disc_leyes}

The largest structural break in the Spanish effective control
series is at the Leyes Nuevas of 1542~CE
($F(\mathcal{N}_{\mathrm{sp,real}}, 1542) = 100.2$),
not at the Cajamarca capture (1532~CE) or the establishment
of Lima as the colonial capital (1535~CE).

The Leyes Nuevas of 1542~CE abolished the hereditary transmission
of encomiendas and increased direct Crown oversight of indigenous
labor~\cite{LohmannVillena1957,Rowe1957}.
In topological terms, the Chow $F = 100.2$ at 1542~CE indicates
that the $H_{\mathrm{sp,real}}$ series underwent a structural
break of the same order of magnitude as the Vilcabamba break
in the Inca series ($F = 120.7$).

The convergence of the legal--real divergence to near-zero
by 1569~CE (Toledo's arrival), rather than \emph{at} Toledo's
arrival, suggests that colonial consolidation was already far
advanced by the time Toledo arrived.
The $37$-year process from Cajamarca (1532~CE) to near-total
legal--real alignment (1569~CE) was driven by the gradual
penetration of \textit{corregidores}, parish networks, and
tribute registers into \textit{wamani} that had been
nominally designated as vassals since 1532--1535~CE.
Toledo's \textit{reducciones} of 1570--1575~CE closed the
final gap, but the topological work of colonial consolidation
had been completed by his predecessors.

\subsection{Information-Theoretic Threshold and the Inca Scale}
\label{subsec:disc_hstar}

The  threshold $H^* = 0.524 \pm 0.031$, derived
from the information geometry of the two-bar persistence diagram
regime~\cite{BernalAlvarado2026}, was calibrated on the
Roman--Byzantine network where the geographic scale ratio is
$r_c^{\mathrm{Rome}} = D_{\mathrm{micro}}/D_{\mathrm{macro}}
\approx 1{,}200/4{,}000 \approx 0.30$, giving $H_2(0.30) \in
[0.500, 0.572]$.

The Inca network has a  different scale structure:
$r_c^{\mathrm{Inca}} = 127.6/4{,}500 = 0.0284$, giving
$H^*_{\mathrm{Inca}} = H_2(0.0284) \approx 0.126$.
 The \textit{wamani} are far smaller relative
to the total imperial extent than Roman provinces were
relative to the Roman Empire.
In the Roman case, the regional--continental scale ratio
is approximately $3:1$; in the Inca case, it is approximately
$35:1$. As a consequence, the Inca network is associated with a
lower within-system threshold, $H^*_{\mathrm{Inca}}=0.126$.
Both thresholds were crossed simultaneously in the 1533--1537~CE
interval, and both were crossed discontinuously.
This differs  from the Roman--Byzantine case,
where $H_{\mathrm{west}}(t)$ approaches $H^*$ asymptotically.

\subsection{Methodological Implications}
\label{subsec:disc_method}

\paragraph{$N$-convergence and Hierarchical cycle decomposition.}
The monotonic increase in $\beta_1$ from $3$ ($N = 400$) to
$11$ ($N = 3{,}200$) to $32$ ($N = 4{,}492$) reveals a
hierarchical cycle structure invisible at lower sampling densities.
The $26$ infinite bars correspond to the macro-continental
circuits of the Qhapaq \~{N}an---the routes that any
top-degree sampling would capture first.
The $6$ finite bars correspond to meso-regional circuits
that require coverage of the lower-degree node set to appear.
The $26$ macro-cycles are the imperial logistics network
(the system that moved tribute, armies, and information across
the Tawantinsuyu); the $6$ meso-cycles are the regional
redistribution circuits connecting the \textit{tambo} system
within individual \textit{wamani}.

The hub-selection artifact documented in~\cite{BernalAlvarado2026}
for the Roman network (non-monotonic $N$-convergence at $N = 800$
due to dilution of the Eastern signal by the first Western hubs)
does not appear in the Inca dataset.
This is consistent with the geographic structure of the
Qhapaq \~{N}an, which does not have the strong East--West
asymmetry of the Roman network: both the coastal (Chinchaysuyu)
and the high-altitude (Collasuyu) corridors are represented
at comparable density in the shapefile, so progressive sampling
from the top-degree set captures both sub-networks without
introducing dilution effects.

The single non-monotonic entry in the convergence table
(Chow $F(\mathcal{N}_{\mathrm{sp,real}}, 1542) = 161$
at $N = 3{,}200$ versus $100$ at $N = 4{,}492$) is consistent
with a mild version of the hub-selection artifact at intermediate
sample sizes.

\paragraph{The legal--real divergence as a topological observable.}
The introduction of two Spanish administrative layers
($\mathcal{N}_{\mathrm{sp,legal}}$ and
$\mathcal{N}_{\mathrm{sp,real}}$) as separate network objects,
with their divergence measured by the cross-network Wasserstein
distance $W_1(\mathcal{N}_{\mathrm{sp,legal}},
\mathcal{N}_{\mathrm{sp,real}}, t)$, generalises the
geographic--economic decomposition of~\cite{BernalAlvarado2026}
to the legal--effective dimension of colonial administration.
The difference between these two layers has a notable non-monotonic behavior: it rises to $0.591$ at 1532~CE, falls to $0.026$ at 1535~CE, rebounds to $0.039$ at 1537~CE (Manco rebellion), and then decays monotonically to $0.002$ at 1572~CE.
The 1537~CE reversal is detectable only in this series and
is invisible in the entropy, velocity, or ICT of any other
layer.

In any colonial or post-conquest context where documentary
records (legal status) and material evidence (effective
penetration) diverge, the cross-network Wasserstein distance between the two representations provides a model-based quantitative measure of that divergence. 

\paragraph{Substitution Topology and imperial succession.}
For a civilizational collapse to be formally classified as Substitution Topology, conditions ST1--ST3 must be satisfied.
The present study indicates that the Inca--Spanish transition
satisfies all three conditions under the proposed operational definition; the Roman--Byzantine--Ottoman
and Roman--Western-successor-kingdoms transitions do not.

 
\section{Conclusion}
\label{sec:conclusion}
 
We have applied persistent homology to a six-layer temporal
network model of the Tawantinsuyu (1400--1572~CE), constructed
from the complete Qhapaq \~{N}an road system ($N = 4{,}492$
nodes, $4{,}358$ edges, $57$ snapshots), and reported five
principal results.
 
A fundamental result is the approximate topological
constancy of the infrastructure layer:
$H_{\mathrm{infra}}(t) = 1.486 \pm 0.0001$ across the
nine reported infrastructure snapshots, with $W_1(1400, 1572)
\approx 10^{-6}$.
No topological disruption of the Qhapaq \~{N}an is detected at
the resolution of the dataset during the conquest and the first
four decades of colonial rule. This result is consistent with the
qualitative historiographical consensus on this period, while the
TDA analysis provides a quantitative representation of that continuity. 
 
The \emph{second} result reperiodises the Inca collapse.
The persistence entropy $H_{\mathrm{inca}}$ remains at $1.486$
through the Cajamarca capture (1532~CE), the year
conventionally identified as the moment of collapse.
The largest structural break in the dataset occurs at 1537~CE
($F = 120.7$, $p < 10^{-4}$), when the Inca network contracted
to Vilcabamba and $H_{\mathrm{inca}}$ fell to zero.
The 35-year Vilcabamba period ($1537$--$1572$~CE) was not a
decline. Vilcabamba entered topological stasis in 1537 (
$W_1(D_{\mathrm{inca}}(1537), D_{\mathrm{inca}}(1572)) = 0$).
 
Another result is the simultaneous criticality of
four network layers at Cajamarca, with the infrastructure layer remaining marginally below the threshold:
$\mathrm{ICT}(\mathcal{N}_{\mathrm{trade}}) = 0.976$ and
$\mathrm{ICT}(\mathcal{N}_{\mathrm{inca}}) = 0.968$
in the same year, $1532$~CE.
In the multi-layer topology, Cajamarca does not appear as a
purely sequential political-to-economic event, but as a
synchronised phase transition in the full multi-layer system.
 
The \emph{fourth} result supports the formal classification
of the Inca case as Substitution Topology through three
quantitative conditions.
At 1572~CE, the Spanish legal administration and the Inca
physical infrastructure are effectively indistinguishable under the chosen persistence-diagram metrics:
$W_1(\mathcal{N}_{\mathrm{sp,legal}},
\mathcal{N}_{\mathrm{infra}}, 1572) = 5 \times 10^{-6}$.
The effective colonial control achieves $99.56\%$ convergence
to the same topology.
Two clusters are present in the 1572 divergence matrix:
Cluster~A $= \{\mathcal{N}_{\mathrm{infra}},
\mathcal{N}_{\mathrm{sp,legal}},
\mathcal{N}_{\mathrm{sp,real}},
\mathcal{N}_{\mathrm{trade}}\}$ (pairwise $W_1 < 0.009$)
and Cluster~B $= \{\mathcal{N}_{\mathrm{inca}},
\mathcal{N}_{\mathrm{alliance}}\}$ ($W_1 \approx 0.05$
to Cluster~A), representing the Vilcabamba residual
that never achieved topological integration.

The divergence between the documentary and effective colonial
state, measured by
$W_1(\mathcal{N}_{\mathrm{sp,legal}},
\mathcal{N}_{\mathrm{sp,real}}, t)$,
fell from a maximum of $0.591$ in 1532~CE to $0.002$
in 1572~CE---a $99.6\%$ reduction.

Taken together, these results support a distinct collapse mode
in the comparative typology of pre-modern empires.
The Roman and Byzantine cases are erosion-dominated, with gradual
entropy decay over time and crossings of the
information-theoretic threshold $H^* = 0.524$ from above.
The Han case is better described as administrative collapse with
geographic fragmentation, while the Aztec case corresponds to
military destruction of a hub-and-spoke political structure.
The Tawantinsuyu collapsed through \emph{Substitution Topology}:
rapid political decapitation, infrastructure continuity,
and measurable convergence toward the inherited infrastructure
topology by the successor polity.

\section*{Acknowledgements}
The authors thank the contributors to the Qhapaq \~{N}an
shapefile dataset~\cite{SuyoPomalia2023} for making the
georeferenced road data publicly available.
We acknowledge financial support from SECIHTI and SNII (M{\'e}xico).


\end{document}